\documentclass{aip-cp}
\usepackage[numbers]{natbib}
\usepackage{rotating}
\usepackage{graphicx}
\begin{document}

\title{Common Origin of Kinetic Scale Turbulence and the Electron Halo in the Solar Wind -- Connection to Nanoflares}

\author{Haihong Che}

\affil[aff1]{University of Maryland, College Park, MD, 20742}
\affil[aff2]{Goddard Space Flight Center, NASA, Greenbelt, MD, 20771}

\maketitle

\begin{abstract}
We summarize our recent studies on the origin of solar wind kinetic scale turbulence and electron halo in the electron velocity distribution function. Increasing observations of nanoflares and microscopic type III radio bursts strongly suggest that nanoflares and accelerated electron beams are common in the corona. Based on particle-in-cell simulations, we show that both the core-halo feature and kinetic scale turbulence observed in the solar wind can be produced by the nonlinear evolution of electron two-stream instability driven by nanoflare accelerated electron beams. The energy exchange between waves and particles reaches equilibrium in the inner corona and the key features of the turbulence and velocity distribution are preserved as the solar wind escapes into interplanetary space along open magnetic field lines. Observational tests of the model and future theoretical work are discussed.
\end{abstract}

\section{Introduction}
There are two long standing puzzles regarding kinetic properties of the solar wind: 1) the origin and nature of kinetic scale turbulence, and 2) the origin of a nearly isotropic electron halo in the electron velocity distribution function (VDF). Observations from 0.3-1 AU show that kinetic scale turbulence and electron halo are prevalent in the solar wind, implying these properties probably have their origins very close to the surface of the Sun. The two puzzles are thought to be closely related to the heating and acceleration of the solar wind\cite{scudder92apja,mak97aap}. 

Observations of solar wind turbulence have shown that as scales approaching the ion inertial length where wave-particle interactions become important, the power-spectrum of magnetic fluctuations, which in the inertial range follows the Kolmogorov scaling $\propto k^{-5/3}$, is replaced by a steeper anisotropic scaling law $B^2_{k_\perp} \propto k_{\perp}^{-\alpha}$, where $\alpha > 5/3$. Spectral index  $\alpha\sim 2.7$ is found in observations but can vary between 2 and 4. Magnetic fluctuations with frequencies much smaller than ion gyro-frequency propagating nearly perpendicularly to the solar wind magnetic field are identified as  kinetic Alfv\'en waves (KAWs) \cite{sah09prl,salem12apjl} and the break frequencies of the magnetic power-spectra suggest the break likely corresponds to the ion inertial length \cite{bou12apj}.  In the past decades, extensive studies of solar wind kinetic scale turbulence have focused on the idea that solar wind kinetic turbulence is due to cascade of large-scale turbulence. However, there are concerns that the energy in the solar wind large-scale turbulence may not be enough to cascade to support the observed kinetic scale turbulence and heating (e.g. \cite{Leamon_et_al_1999}). 

Observations of electron VDFs at heliocentric distances from 0.3 to 1 AU show a prominent ``break" or a sudden change of slope at a kinetic energy of a few tens of electron volts. The electron VDF below the break is dominated by a Maxwellian known as the ``core" while the flatter wing above the break is called the ``halo"\cite{pilipp87jgra,pilipp87jgrb}. So far no model can naturally produce the nearly isotropic halo population which can be described by a kappa function\cite{marsch06lrsp}. The isotropic nature of the halo suggests that halo formation needs strong turbulence scattering and is likely related to the kinetic turbulence in the solar wind\cite{marsch06lrsp}. In addition, ``strahl" -- an anisotropic tail-like feature skewed with respect to the magnetic field direction is found in the electron VDF of solar wind with speed $>\sim 400$~km~s$^{-1}$.  In the solar wind coming from the sector boundary with speed $\sim 400$~km~s$^{-1}$, the strahl is nearly invisible and the isotropic core-halo feature dominates. Existing kinetic models based on magnetic focusing effect with input of different modes of kinetic turbulence, while successful in producing the strahl-like tail at a minimum heliocentric distance of 10 $R_{\odot}$\cite{smith_hm12apj,landi12apja}, are unable to produce the halo\cite{marsch06lrsp,marsch12ssr,vocks12ssr}. This suggests that extra energy dissipation, probably by plasma kinetic instabilities, is required \cite{marsch06lrsp,marsch12ssr}. 
At large heliocentric distances (0.3 - 1 AU), observations found the electrons are scattered from strahl\cite{mak05jgr,stverak09jgr,gurg12ag}, and theories suggest that the scattering might be caused by the unstable processes related to the strahl\cite{vocks05apj,mith12pop}. 

Clearly to generate the observed isotropic halo and possibly the observed turbulent fluctuations on kinetic scales,  some form of  ``extra free energy" and instabilities on kinetic scales are needed. What is the source of this free energy?

\section{Nanoflares and a Unified Model for the Origin of Kinetic Turbulence and The Electron Halo}

High spatial and spectral resolution observations of the Sun from SOHO/SUMER and TRACE together with Extreme Ultraviolet Imaging Telescope (EIT) have revolutionized our view of the origin of solar wind. Different from the steady fluid solar wind model\cite{parker58apj}, new observations\cite{feldman05jgr,marsch07esa} found that the solar wind originates from impulsive events close to the surface of the Sun. The correlation between the Doppler-velocity and maps of radiance of spectral lines emitted by ions of various charge states in the solar atmosphere suggests that the solar wind originates from small magnetic loops rooted in the photosphere and escapes along open field lines caused by the merging of loops through granular convection\cite{xia03aap, feldman05jgr,tu05sci}. 

The emerging dynamic picture of solar wind strongly suggests nanoflares \footnote{By nanoflare we refer to small scale explosive events that occur everywhere in the quiet Sun, including corona holes. The estimated occurrence rate of such events is $\sim 10^6$~$\rm{s}^{-1}$ for the whole Sun. Our use of the term is an extension of Parker's original definition\cite{parker88apj}} contribute significantly to the solar wind. Recent high resolution observation of the Sun from rockets and IRIS are providing increasing detailed pictures of nanoflares\cite{win13apj,testa14sci}. Similar to flares, nanoflares can accelerate particles and the characteristic energy of nanoflare-accelerated electrons is in keV range\cite{gon13apj}. The accelerated electron beams can trigger the electron two-stream instability (ETSI), generate Langmuir waves, and produce type III radio bursts. Indeed, observations\cite{saint13apj} have found in the solar corona a new type of radio III burst whose brightness temperatures are about 9 orders of magnitude lower than flare-associated Type III bursts and are far more abundant, implying the bursts very possibly originate from nanoflares. The high occurrence rate of these ``nano type III bursts" indicates that electron streams and streaming instabilities are common in the solar corona, and the instabilities can contribute to the kinetic properties of the solar wind. The energetic charged particle streams produced by nanoflares and subsequent streaming instabilities are very likely source of the free energy that generates both the solar wind kinetic turbulence and the isotropic electron halo\cite{che14apjl}.  

How does the ETSI driven by nanoflares shape the kinetic properties of the solar wind? 
In two recent papers\cite[]{che14prl,che14apjl}, we showed that the nanoflare-driven ETSI can produce the observed kinetic turbulence and electron halo can form in about 10 gyro-periods $\Omega_i t\sim 10$ or $\omega_{pe} t\sim 10 c/v_A \sqrt{m_i/m_e} $ electron plasma oscillation periods via nonlinear evolution of ETSI in the solar inner corona. The coexistence of wave turbulence and electron halo is a turbulence equilibrium stage in which the energy loss and gain between waves and particles are statistically equal. The heated particles are diffused into the solar corona from the magnetic loops. Both electric and magnetic fluctuations and the non-Maxwellian VDFs are advected away from the corona with the wind along open magnetic field lines produced by the reconnection of loops due to the convective granulation of plasma in photosphere\cite{gloeckler03jgr,fisk03jgr}. Because in our model kinetic turbulence is generated in the corona, properties of the solar wind can be affected by local kinetic processes when the wind travels to 1AU and beyond. However, the kinetic turbulence advected from the solar corona is less affected in the slow wind (with velocity $<$ 400~km~s$^{-1}$) since the electron VDF of slow wind is nearly isotropic and hence is least prone to unstable processes that can be triggered by the anisotropic electron VDF of fast wind ($>500$~km~s$^{-1}$)\cite{che14apjl}. 

PIC simulations have been performed to investigate the nonlinear evolution of ETSI. The code is a massive parallel explicit PIC code. The simulations fully resolve the physical processes spanning from Debye length to ion inertial length and electron plasma frequency to ion cyclotron frequency. The simulations use high spatial and time resolution that effectively reduce the simulation noise, in particular, numerical heating on electron scales. The simulation box has dimensions of  $32 \times 32 d_i^2$, and the particle number is up to $10^{11}$. The mass ratio is $m_i/m_e=100$ and the speed ratio of light and Alfv\'en wave is $c/v_A=100$. $\beta=0.25$ and electron and ion temperature is equal. The density of electron beam is $10$\%  of the core electrons. The total grid number is $10420 \times 10420$ and the particle number per cell is 100.
\begin{figure}
\includegraphics[scale=1.15, trim = 70 540 80 70,clip]{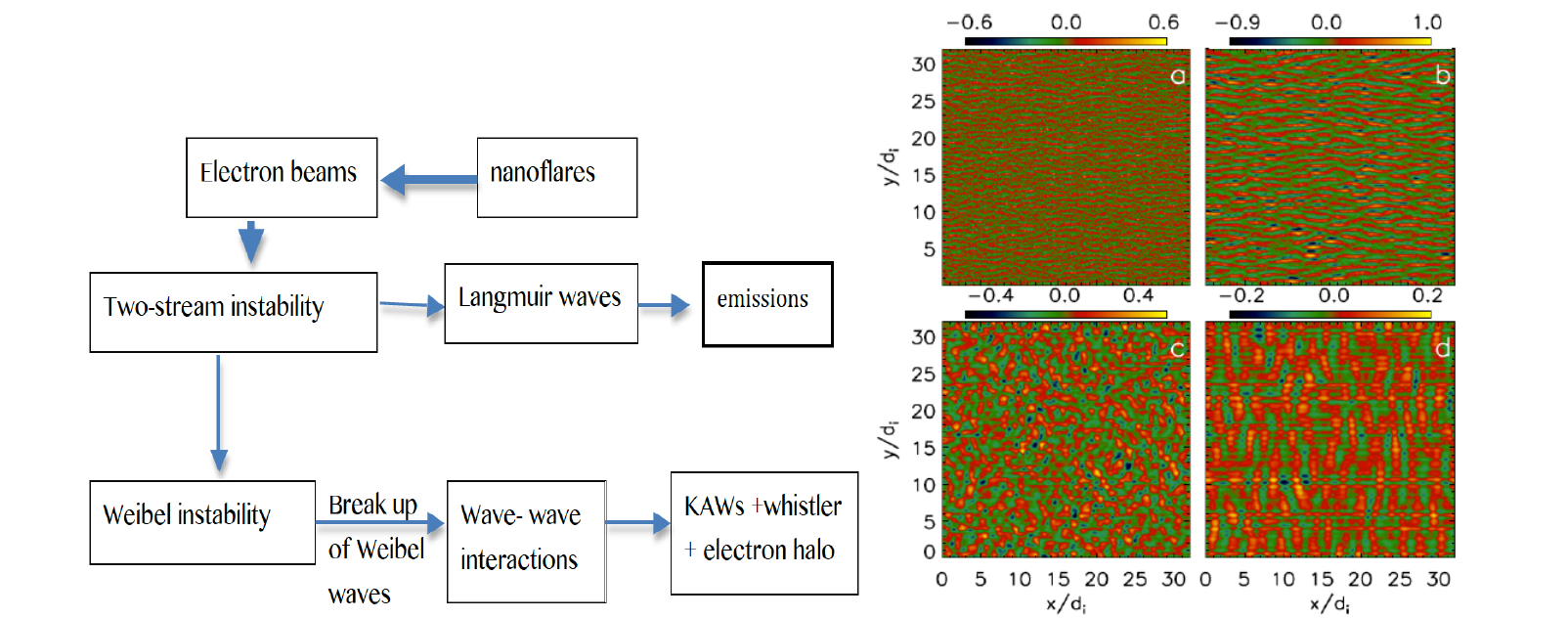} 
\caption{ {\bf Left Panel} The flowchart showing the generation of waves during the nonlinear evolution of ETSI. {\bf Right panel} Images of $B_z/B_0$ at $\omega_{pe} t=$ 24, (panel {\bf a}) fast growth of electric field induced magnetic field, $\omega_{pe} t=$ 480 (panel {\bf b}) Weibel-like instability is triggered, $\omega_{pe} t=$ 2424 (panel {\bf c}) irregular wave-wave interactions, and $\omega_{pe} t=$ 10560 (panel {\bf d}) parallel propagating whistler wave and perpendicular propagating KAWs are produced.  }
\label{mychart}
\end{figure}

Simulations show that the evolution of ETSI have four distinctive stages (Fig. 1) : (a) ETSI rapidly (on electron plasma oscillation time-scale $\sim 1/\omega_{pe}$) converts $\sim 90\%$ of the kinetic energy of the electron stream into the heat of ambient electrons and converts $\sim 10\%$ into magnetic energy due to the fast growth of electric field, and then gradually evolves to full saturation; (b) At $\omega_{pe} t \sim  200 $, ETSI generates electron holes which trap electrons and cause localized currents. The currents in turn generate an electromagnetic Weibel-like instability which reaches its peak at $\omega_{pe} t \sim  500 $. This is the stage when electrostatic wave is converted to electromagnetic waves; (c) The Weibel-like electromagnetic waves break up into small pieces, and propagate randomly (Panel (c) in Fig.1),
accompanied by strong wave-wave interactions at $\omega_{pe} t \sim 2000$. (d) Eventually at  $\omega_{pe}t \sim 10000$ waves interact to produce both perpendicular propagating KAWs and parallel propagating whistler waves relative to the background magnetic field and the turbulence fully saturates, i.e. the energy exchanges between waves and particles reaches balance. Inverse energy cascade produces KAWs and stops at inertial length while forward energy cascade produces whistler waves that starts at ion gyro-radius. The wave scattering leads to a nearly isotropic hot electron tail, i.e. the electron halo. The simulation show that Langmuir waves and emissions with electron plasma frequency are generated in Stage (b) and continue to grow beyond stage (d) when the kinetic turbulence reaches turbulent equilibrium with the electron halo(Che et al. 2015, to be submitted to {\textit{Proceedings of the National Academy Science of the United States of America}).

The model produces properties of the electron halo \cite{che14apjl} and the power-spectrum of KAWs  which are consistent with observations, and the model predicts the existence of both KAWs and whistler waves in the solar wind turbulence. In addition, production of Langmuir waves and plasma emissions in our simulation lends support for the nanoflare origin of the low brightness temperature Type III radio bursts. Enhanced electrostatic fluctuations are seen in our simulations, which could explain what has been observed in \cite{mozer13apjl}.  We briefly summarize these results below: 

{\bf  Turbulence on kinetic scales: } In Fig.2, we show the simulated power spectra of $(\delta B)^2$ in directions both parallel and perpendicular to the background magnetic field (In our simulation the background magnetic field is in x-direction). With $1<k_y d_i<2$, which corresponds to the range of wave lengths current instruments can probe, both KAWs and whistler waves are important. The perpendicular power spectrum is fitted with a power-law with an index of -2.2. The power-law indices from solar wind observations have large variations, and the spectral index from our simulation falls within the observed range. The perpendicular power spectrum terminates at the ion inertial length, also consistent with observations\cite{Leamon_et_al_1999,perri10apjl,bou12apj}. A unique feature of our model is the spectral break at the electron scale caused by energy injection. This model also predicts the existence of whistler waves,  and the parallel power spectrum terminates at the ion gyro-radius\cite[]{che14prl}. 

\begin{figure}
\includegraphics[scale=1.1, trim = 70 580 70 50,clip]{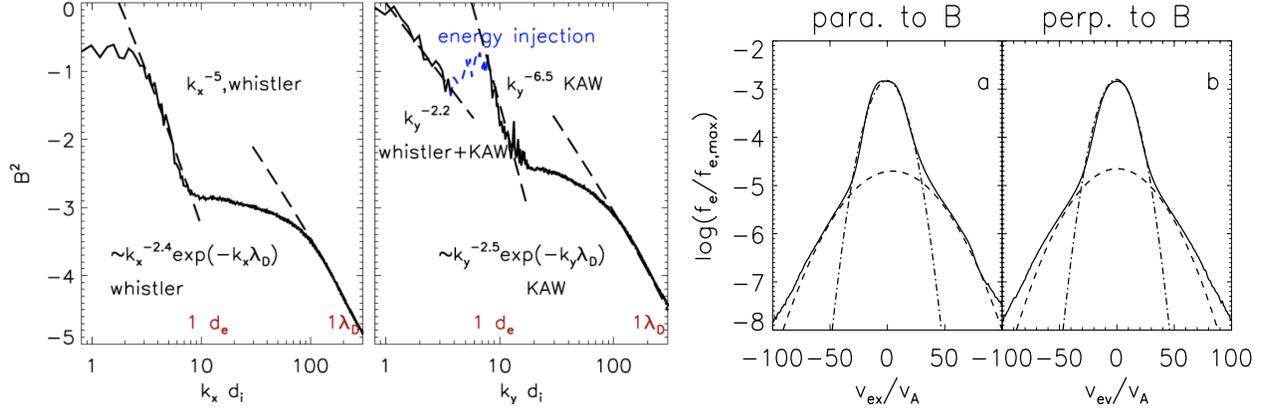} 
\caption{{\bf Left Panel} The two 1D spectra of $ \delta B^2(k)$ vs. $k_x d_i$ (parallel to $B_0$) and  $k_y d_i$ (perpendicular to $B_0$). The blue short-dashed line is used to highlight  the  energy injection that breaks the continuity of the spectrum and produces random fluctuations.  The coordinate system is the same as in Fig.~1. {\bf Right Panel} The 1D electron VDFs  $f(v_x,v_y=0)$ (parallel to $B_0$) and $f(v_x=0, v_y)$ (perpendicular to $B_0$) at $\omega_{pe} t=10560$. The dot-dashed lines delineate the core Maxwellian distribution functions with $T_{c}=1 m_i v_A^2$ and the dashed lines represent the halo distribution functions with $T_{h}\sim 6 m_i v_A^2$.}
\label{spec1d}
\vspace{10pt}
\end{figure}
{\bf  Core-halo structure of the electron VDF:} A major result of the model is the core-halo structure of solar wind electron VDF which  forms in the inner corona and the solar wind advects the features into interplanetary medium along open field lines. In Fig.~\ref{spec1d} we show two simulated electron VDFs which are plotted in the same manner as in Pilipp et al. (1987). The temperatures of core and halo are defined by their corresponding best-fit Maxwellians. Our model predicts that the core-halo temperature ratio $T_h/T_c$ of the solar wind is insensitive to the initial conditions in the corona and is related to the core-halo density ratio of the solar wind $n_c/n_h$:
\begin{equation}
\frac{T_h}{T_c}  \approx \frac{n_c}{n_h}\frac{1-C_T}{C_T}+4,
\label{dthc2}
\end{equation} 
where $C_T$ is the rate at which kinetic energy of electron beams converts to heat, and $C_T \sim 0.9$ in our simulation. This relation can be extended to the more general core-halo-strahl feature in the solar wind because the energy is conserved during the formation of strahl caused by magnetic focusing effect:
\begin{equation}
\frac{T_{hot}}{T_c} \sim \frac{(1-C_T)}{C_T}\frac{n_c}{ n_{hot}}+4,
\end{equation}
where $T_{hot}=(n_h T_h + n_s T_s)/(n_h+n_s)$ is the mean hot component temperature, and subscript $s$ represents the strahl.
If we assume the core and halo experience similar temperature evolutions when traveling from the Sun to 1AU, the temperature ratio can be approximately preserved. 

The break point dividing the core and halo in electron VDF, which is a useful quantity in observations, satisfies:
\begin{equation}
v_{brk}\approx[\ln(T_{h}/T_c)-\ln (n_h/n_c)^2]^{1/2} v_{te,c}.
\label{brk}
\end{equation} 
In addition, the relative drift between the core and halo is close to the core thermal velocity -- a relic of ETSI saturation.

{\bf  Langmuir Waves and Type III Radio Bursts Associated with Nanoflares:}
Our simulations show that both Langmuir waves and electromagnetic emissions are produced during the development of ETSI. In the solar corona these high frequency electromagnetic emissions are in the radio band. Since the density of corona decreases with the radial distance from the Sun, the corresponding wave frequency decreases too when Langmuir waves propagate outward in the corona, resulting in the frequency drift, characteristic of Type III radio bursts. These are qualitatively consistent with the expectation that the ``nano Type III radio bursts" occurring in the solar corona with very low temperature brightness $\sim 10^6-10^8$ K\cite[]{saint13apj}  compared to $10^{15}$ K for normal Type IIIs are associated with flares accelerated electron beams. 

{\bf  Enhanced electrostatic fluctuations on kinetic scales:} Our simulations\cite{che14prl} show that when the kinetic turbulence fully saturates, the ratio of parallel to perpendicular electric field fluctuations $\langle \vert \delta E_{\parallel}\vert/\vert \delta E_{\perp}\vert\rangle$ is enhanced by the relic parallel electric field by a factor of $\sim 2-3$, consistent with observations that the parallel turbulent electric field is larger than the perpendicular turbulent electric field, contrary to what is expected if the turbulent fluctuations are dominated by KAWs \cite{mozer13apjl}. The enhanced electric field might be caused by electrostatic whistler wave and Langmuir wave. 

\section{Conclusions and Future Work}
The nanoflare-accelerated electron beams are the possible ``free energy" source responsible for the generation of kinetic turbulence and the electron halo observed in the solar wind. As we have shown the model can produce existing observed solar wind kinetic properties and makes predictions that are testable by utilizing archival solar wind data as well as data from future space missions. One major attraction of our model is that it solves both puzzles of the electron VDF and kinetic turbulence in a unified manner, while past studies of these two  phenomena treat them as unrelated.  The link between the solar wind and nanoflares directly relates solar wind properties to photosphere dynamics, putting useful constraints on kinetic processes in both solar corona and the solar wind.  

While our model predictions are consistent with existing observations, there are predictions unique to this model and can be observationally tested: 

1) In our model, the electron halo forms in the inner corona. On the other hand, if the electron halo is produced by the scattering of strahl electrons as some models suggest, the halo would be much weaker at heliocentric distances $\sim 10$R$_{\odot}$ than our model predicts. This is because it takes at least  $\sim 10$R$_{\odot}$ for the strahl to develop and become unstable by magnetic focusing effect to produce electron scattering\cite{smith_hm12apj}. Future mission such as {\it Solar Probe Plus} (SPP)  will be able to observe electron VDF at 10 R$_{\odot}$ and settle the issue. 

2) The energy injection on electron scale produces a plateau in the magnetic fluctuation power spectrum, and parallel whistler wave turbulence terminates at the ion gyro-radius, in contrast to KAW turbulence which terminates at ion inertial length. Such features may only be testable at small heliocentric distances because at 1 AU the spectrum might be affected by the various events when the solar wind travels in the interplanetary space. In-situ observations of SPP at 10 R$_{\odot}$ should be more suitable for detecting the plateau and spectral breaks. 

3) Parallel propagating whistler waves, Langmuir wave and the subsequent enhanced parallel electric field are features of our model. In particular, parallel whistler wave is a sign if the energy exchange between wave and electron halo reaches equilibrium. If the electron halo becomes anisotropic along magnetic field line (strahl), the oblique whistler wave will be generated which produces electron scattering until the electron halo becomes isotropic again. Thus in the solar wind, it is more likely that both the parallel and oblique whistler waves can be observed. Such features could be observable by current and future space observations.  

4) The nanoflare-produced type III bursts are about 9 orders of magnitudes weaker than the flare associated type III bursts. Both the SPP and {\it Solar Orbiter} (Zouganelis, in these proceedings) plasma wave detectors have the capability to observe these bursts (Maksimovic and Bale, private communications). Electron beams can also affect Balmer line emission and produce X-ray and EUV emissions. 

In addition, nanoflares are also likely to be responsible for the so called ``super-halo" observed in the solar wind electron VDF by WIND and STEREO space crafts\cite{wang12apjl}. Such feature can be a natural consequence of the energy distribution function of the nanoflare accelerated electron beams which should have a high energy tail. The ``tail electrons" could be scattered by the kinetic waves generated by the bulk of the beam electrons to produce the observed super-halo. Such a possibility needs further investigation with large PIC simulations.  

So far we have only studied the electron dynamics in the solar wind. Since nanoflares also accelerate ions, and ion dynamics is critically important to the formation and accelerations of solar wind. This requires us to incorporate ion beams into our model to understand the dynamics of ions as well as how ion and electron waves interact. 

Our current simulations are 2.5D since the system we are studying is symmetric in the third dimension. We will investigate 3D effects in the near future to study if the waves propagating perpendicular to magnetic field develop in the third dimensions and enhance the wave-wave and wave-particle interactions. 

\section{ACKNOWLEDGMENTS}
This research was partially supported by the NASA Postdoctoral Program and NASA research grant. The work is in collaboration with Drs. M. L. Goldstein and A. F. Vi\~nas. The simulations and analysis were carried out at the NASA Advanced Supercomputing (NAS) facility at the NASA Ames Research Center. 


\bibliographystyle{aipproc}
\bibliography{sw14}%

\end{document}